\documentclass[a4paper,aps,onecolumn,nofootinbib]{revtex4}
\RequirePackage[colorlinks,hyperindex]{hyperref}
\RequirePackage[english]{babel}
\RequirePackage[latin1]{inputenc}
\RequirePackage[T1]{fontenc}
\RequirePackage{mathrsfs}
\RequirePackage{amsmath}
\RequirePackage{amssymb}
\RequirePackage{amsbsy}
\RequirePackage{color}
\RequirePackage{bm}
\usepackage{graphicx}
\usepackage{float}
\hypersetup{colorlinks=true,breaklinks=true,urlcolor=blue,linkcolor=red}
\begin{document}
\title{\bf{General Spin Sums in Quantum Field Theory}}
\author{Rodolfo Jos\'{e} Bueno Rogerio$^{\hbar}$\footnote{rodolforogerio@gmail.com} and
Luca Fabbri$^{\nabla}$\footnote{luca.fabbri@edu.unige.it}}
\affiliation{$^{\hbar}$Instituto de F\'isica e Qu\'imica,\\
Universidade Federal de Itajub\'a - IFQ/UNIFEI,\\
Av. BPS 1303, CEP 37500-903, Itajub\'a - MG, BRAZIL\\
$^{\nabla}$DIME, Sez. Metodi e Modelli Matematici,\\
Universit\`{a} di Genova,\\
Via all'Opera Pia 15, 16145 Genova, ITALY}
\date{\today}
\begin{abstract}
In Quantum Field Theory, scattering amplitudes are computed from propagators which, for internal lines, are built upon spin/polarization-sum relationships. In turn, these are normally constructed upon plane-wave solutions of the free field equations. A question that may now arise is whether such spin/polarization-sums can be generalized. In the past, there has been a first attempt at generalizing spin sums for fermionic fields in terms of the Michel-Wightman identities. In this paper, we aim to find the most general spin sums for fermionic fields within the range of QFT.
\end{abstract}
\maketitle
\section{Introduction}
Quantum Field Theory is the most important tool we have to compute scattering processes in particle physics. The actual calculations of scattering amplitudes are based on perturbative expansions that reduce the interaction of the scattered particles to combinations of freely-propagating particles and point-like processes in which all information about the interaction is encoded. These freely-propagating particles can be further distinguished according to whether they are the observable particles coming/going from/to infinity or the virtual particles that appear as internal lines between vertices. Virtual particles describing internal lines are themselves represented by propagators, written in terms of spin-sum and polarization-sum relationships. Therefore these spin sums and polarization sums are the fundamental quantities from which propagators can be evaluated, and scattering amplitudes can be computed.

As is normal in QFT, spin sums and polarization sums are commonly calculated from fields that are taken in the form of plane-wave solutions for the free field equations. Nevertheless, one may wonder what would happen when more general solutions are considered. A first extension was made by Michel and Wightman who found the spin sums of fermion fields in cases in which the spin axial-vector of the spinor was not to be averaged out \cite{mw}. The Michel-Wightman spin sums display a corresponding enlarged structure. In the present work, our goal is to continue such an enlargement up to include spin sums for fermions in the most general case allowed in standard QFT.
\section{Recalling the Fundamentals: Particle Scattering}
\label{II}
We begin our presentation by specifying that all our results are found within the formalism of standard Quantum Field Theory, and that all generalizations are only about the content of the spin sums. As for the QFT notations, we follow those of reference \cite{ps}. This means that, in particular, spinors are promoted to operators, themselves verifying anti-commutation relationships, as usually done in QFT. Notice that all results about locality and causality, and in particular the fact that the anti-commutators vanish for space-like separation, are consequently ensured by the very construction of the theory (for more details we direct the reader to \cite{ps} chapter $3$).

Additionally, all scattering amplitudes are considered to occur, to a very good approximation, only between point-like particles, with no internal dynamics nor extension. For asymptotic states given by an initial $|\psi\rangle_{i}$ and a final $|\psi\rangle_{f}$ the amplitude of the initial state to scatter off to the final state is given by
\begin{eqnarray}
S_{fi}\!=\!_{f}\langle\overline{\psi}|U|\psi\rangle_{i}\label{S}
\end{eqnarray}
where
\begin{eqnarray}
U\!=\!\exp{\left(-i\int_{-\infty}^{+\infty}\!\!\!\! H_{\mathrm{int}} dt\right)}\label{U} 
\end{eqnarray}
in which $H_{\mathrm{int}}$ is the Hamiltonian of the interaction. Since $U$ is an exponential, we know its perturbative series and therefore we can write explicitly all terms in powers of the Hamiltonian of interaction (see \cite{ps} chapter $4$).

For Quantum Electrodynamics, the Hamiltonian is given by
\begin{eqnarray}
H_{\mathrm{int}}\!=\!\int_{V} e \overline{\psi}\boldsymbol{\gamma}^{\mu}\psi A_{\mu}dV\label{H}
\end{eqnarray}
in which $\boldsymbol{\gamma}^{\mu}$ is an element of the Clifford algebra needed to define the spinor $\psi$ with charge $e$ and $A_{\mu}$ is the electrodynamic field. Thus in QED the elementary process has Feynman diagram
\begin{figure}[H]
\includegraphics[bb=40 650 600 770,scale=0.7]{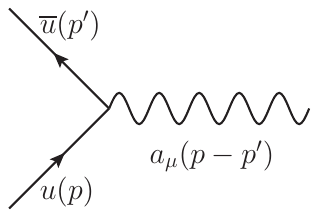}
\label{fig-4}
\end{figure}
\noindent so that every internal line is obtained as a hinge between two external legs having equal momenta. In this case, the two hinged lines constitute either a polarization sum or a spin sum that can be evaluated, for plane waves, to be
\begin{eqnarray}
a_{\mu}(p)a_{\nu}(p)\!=\!\frac{1}{p^{2}}(-g_{\mu\nu})\label{aa}
\end{eqnarray}
and
\begin{eqnarray}
u(p)\overline{u}(p)\!=\!\frac{1}{p^{2}\!-\!m^{2}}(p\!\!\!/\!+\!m\mathbb{I})\label{uu}
\end{eqnarray}
respectively (readers interested in details can find the review of Feynman rules for QED in \cite{ps} chapters $4$ and $5$).

Here we give just a few of the simplest examples. The first is given by the electron-positron scattering
\begin{figure}[H]
\includegraphics[bb=10 550 600 770,scale=0.7]{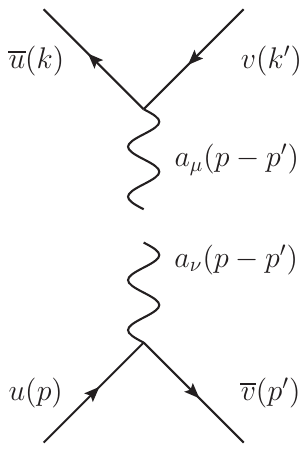}
\label{fig-5}
\end{figure}
\noindent whose invariant matrix element is
\begin{eqnarray}
&\mathcal{M}\!=\!\overline{v}(p')(ie\boldsymbol{\gamma}^{\mu})u(p)
a_{\mu}(p\!-\!p')a_{\nu}(p\!-\!p')\overline{u}(k)(ie\boldsymbol{\gamma}^{\nu})v(k')
\end{eqnarray}
with $a_{\mu}(p\!-\!p')a_{\nu}(p\!-\!p')$ to be evaluated with (\ref{aa}). Therefore the electron-positron scattering is given by
\begin{figure}[H]
\includegraphics[bb=10 570 600 780,scale=0.7]{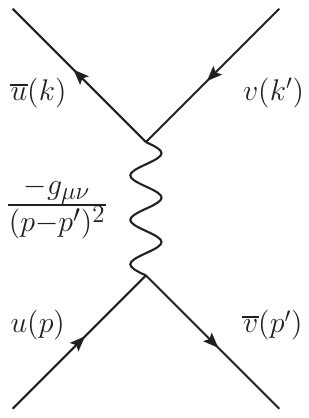}
\label{fig-1}
\end{figure}
\noindent and its invariant matrix element is
\begin{eqnarray}
\mathcal{M}\!=\!\overline{v}(p')(ie\boldsymbol{\gamma}^{\mu})u(p)
\frac{-g_{\mu\nu}}{(p\!-\!p')^{2}}
\overline{u}(k)(ie\boldsymbol{\gamma}^{\nu})v(k')
\label{M1}
\end{eqnarray}
with $u(p)$ plane-wave spinor corresponding to an electron and $v(p')$ plane-wave spinor corresponding to a positron (the reader may check with \cite{ps} page 131). Alternatively, the complementary case is the electron-photon scattering
\begin{figure}[H]
\includegraphics[bb=10 550 600 770,scale=0.7]{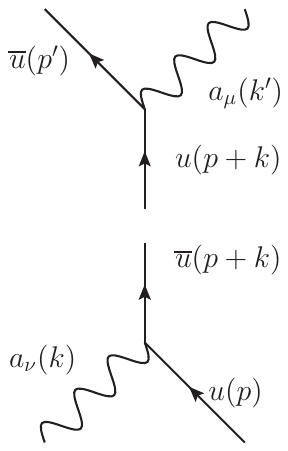}
\label{fig-6}
\end{figure}
\noindent so that the invariant matrix element is
\begin{eqnarray}
&\mathcal{M}\!=\!\overline{u}(p')a_{\mu}(k')(ie\boldsymbol{\gamma}^{\mu})
u(p+k)\overline{u}(p+k)(ie\boldsymbol{\gamma}^{\nu})a_{\nu}(k)u(p)
\end{eqnarray}
in which $u(p+k)\overline{u}(p+k)$ is calculated with (\ref{uu}). Hence the electron-photon scattering is given by
\begin{figure}[H]
\includegraphics[bb=10 570 600 780,scale=0.7]{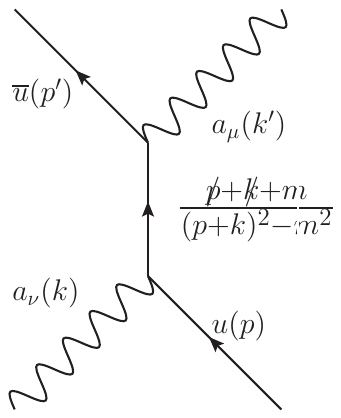}
\label{fig-2}
\end{figure}
\noindent whose invariant matrix element is
\begin{eqnarray}
\mathcal{M}\!=\!\overline{u}(p')a_{\mu}(k')(ie\boldsymbol{\gamma}^{\mu})
\frac{p\!\!\!/\!+\!k\!\!\!/\!+\!m\mathbb{I}}{(p\!+\!k)^{2}\!-\!m^{2}}
(ie\boldsymbol{\gamma}^{\nu})a_{\nu}(k)u(p)
\label{M2}
\end{eqnarray}
where $a_{\rho}$ is the polarization vector of the photon (again see \cite{ps} page 158). In the same way one might re-construct all higher-order combinations, including loop diagrams of radiative corrections (see again \cite{ps} chapters $5$ and $6$).

This perhaps trivial summary of Feynman rules serves the purpose of highlighting how the scattering amplitudes can be computed in terms of internal lines hinged by using polarization-sum and spin-sum relationships. The polarization and spin sums themselves are evaluated on specific solutions of the field equations. Thus the construction that leads to the computation of the matrix elements is based on expressions (\ref{S}-\ref{U}), which are general, and (\ref{H}), which is specific of a given interaction, but for that interaction it is still general. However, expressions (\ref{aa}-\ref{uu}) are valid only for plane waves. Consequently, to consider solutions that are more general than plane waves, the spin sums must be more general than (\ref{aa}-\ref{uu}). The rest of the construction given by (\ref{S}-\ref{U}, \ref{H}) is the same as that of standard QED.

Now, to enlarge the spin sums from the usual (\ref{aa}-\ref{uu}), we may notice that a first step has already been undertaken in \cite{mw} by Michel and Wightman, when they dropped the assumption of averaging out the spin axial-vector, obtaining the more general Michel-Wightman identities. In the following section, we will find a further extension of the spin sums for fermion fields that will generalize the Michel-Wightman spin sums up to their most general case.

To do that, we need a tool that is known in literature as the polar decomposition.
\section{The Polar Form}
\label{III}
The polar decomposition of spinor fields is essentially the technique for which in a spinor each complex component can be written in polar form, that is as a product of a module times a phase, while maintaining the manifest covariance of the formulation. After writing spinors in polar form, it is possible to easily compute the spin-sum relationships \cite{Fabbri:2020ypd}.

To begin, we introduce the Clifford matrices $\boldsymbol{\gamma}^{a}$ verifying the Clifford algebra $\{\boldsymbol{\gamma}^{a},\boldsymbol{\gamma}^{b}\}\!=\!2\mathbb{I}\eta^{ab}$ in which $\eta^{ab}$ is the Minkowski matrix. Then we also have $\boldsymbol{\sigma}^{ab}\!=\![\boldsymbol{\gamma}^{a},\boldsymbol{\gamma}^{b}]/4$ satisfying $2i\boldsymbol{\sigma}_{ab}\!=\!\varepsilon_{abcd}\boldsymbol{\pi}\boldsymbol{\sigma}^{cd}$ in terms of the completely antisymmetric pseudo-tensor $\varepsilon_{abcd}$ and implicitly defining the $\boldsymbol{\pi}$ matrix (this is usually denoted as $\boldsymbol{\gamma}^{5}$ but we prefer to use a notation with no index). Given a spinor $\psi$ and its adjoint $\overline{\psi}\!=\!\psi^{\dagger}\boldsymbol{\gamma}^{0}$ we can build the spinorial bi-linears as
\begin{eqnarray}
&\Sigma^{ab}\!=\!2\overline{\psi}\boldsymbol{\sigma}^{ab}\boldsymbol{\pi}\psi
\ \ \ \ \ \ \ \ S^{a}\!=\!\overline{\psi}\boldsymbol{\gamma}^{a}\boldsymbol{\pi}\psi
\ \ \ \ \ \ \ \ \ \Theta\!=\!i\overline{\psi}\boldsymbol{\pi}\psi\\
&M^{ab}\!=\!2i\overline{\psi}\boldsymbol{\sigma}^{ab}\psi
\ \ \ \ \ \ \ \ U^{a}\!=\!\overline{\psi}\boldsymbol{\gamma}^{a}\psi
\ \ \ \ \ \ \ \ \ \Phi\!=\!\overline{\psi}\psi
\end{eqnarray}
which are all real tensors and such that they verify
\begin{eqnarray}
&\psi\overline{\psi}\!\equiv\!\frac{1}{4}\Phi\mathbb{I}
\!+\!\frac{1}{4}U_{a}\boldsymbol{\gamma}^{a}
\!+\!\frac{i}{8}M_{ab}\boldsymbol{\sigma}^{ab}
\!-\!\frac{1}{8}\Sigma_{ab}\boldsymbol{\sigma}^{ab}\boldsymbol{\pi}
\!-\!\frac{1}{4}S_{a}\boldsymbol{\gamma}^{a}\boldsymbol{\pi}
\!-\!\frac{i}{4}\Theta \boldsymbol{\pi}\label{Fierz}
\end{eqnarray}
called Fierz identities. We can now give a very general result about how spinors can be re-expressed. This result is about the classification of spinor fields, which can be very generally split in two major classes: one is given by those spinors verifying the constraints $\Theta\!=\!\Phi\!\equiv\!0$ called singular spinors; the other is given by the spinors having in general $\Theta^{2}\!+\!\Phi^{2}\!\neq\!0$ called regular spinors. Singular spinors, containing Weyl and Majorana spinors \cite{L, Cavalcanti:2014wia, HoffdaSilva:2017waf, daSilva:2012wp,Cavalcanti:2020obq,Ablamowicz:2014rpa, Rodrigues:2005yz}, are a very important class of spinors, but we are not going to focus on them in this work. We will instead focus on regular spinors. Regular spinors can always be written (in chiral representation) according to
\begin{eqnarray}
&\!\!\psi\!=\!\phi e^{-\frac{i}{2}\beta\boldsymbol{\pi}}
\boldsymbol{L}^{-1}\left(\!\begin{tabular}{c}
$1$\\
$0$\\
$1$\\
$0$
\end{tabular}\!\right)
\label{regular}
\end{eqnarray}
in terms of $\boldsymbol{L}$ being a specific complex Lorentz transformation, and in which $\beta$ and $\phi$ are a real pseudo-scalar and a scalar called chiral angle and module. A straightforward substitution in (\ref{Fierz}) of (\ref{regular}) gives
\begin{eqnarray}
&\psi\overline{\psi}\!\equiv\!\frac{1}{2}\phi^{2}e^{-i\beta\boldsymbol{\pi}}
(e^{i\beta\boldsymbol{\pi}}\!+\!u_{a}\boldsymbol{\gamma}^{a})
(e^{-i\beta\boldsymbol{\pi}}\!-\!s_{a}\boldsymbol{\gamma}^{a}\boldsymbol{\pi})
\label{Fierzpolarregular}
\end{eqnarray}
in the most general case and having defined $S^{a}\!=\!2\phi^{2}s^{a}$ and $U^{a}\!=\!2\phi^{2}u^{a}$ as the spin axial-vector and velocity vector.

With spinors in polar form, the corresponding spinorial covariant derivative can always be written as
\begin{eqnarray}
&\!\!\!\!\!\!\!\!\boldsymbol{\nabla}_{\mu}\psi\!=\!(-\frac{i}{2}\nabla_{\mu}\beta\boldsymbol{\pi}
\!+\!\nabla_{\mu}\ln{\phi}\mathbb{I}
\!-\!iP_{\mu}\mathbb{I}\!-\!\frac{1}{2}R_{ij\mu}\boldsymbol{\sigma}^{ij})\psi
\label{decspinder}
\end{eqnarray}
in terms of $P_{\mu}$ being the usual momentum associated to the gauge structure and where $R_{ij\mu}$ is a tensor playing the role of momentum associated to the space-time structure \cite{Fabbri:2020ypd}. Then, the Dirac spinor field equations in polar form are
\begin{eqnarray}
&\nabla_{\mu}\beta\!+\!B_{\mu}\!-\!2P^{\iota}u_{[\iota}s_{\mu]}
\!+\!2s_{\mu}m\cos{\beta}
\!=\!0\label{dep1}\\
&\nabla_{\mu}\ln{\phi^{2}}\!+\!R_{\mu}
\!-\!2P^{\rho}u^{\nu}s^{\alpha}\varepsilon_{\mu\rho\nu\alpha}
\!+\!2s_{\mu}m\sin{\beta}
\!=\!0\label{dep2}
\end{eqnarray}
where we have introduced $R_{\mu a}^{\phantom{\mu a}a}\!=\!R_{\mu}$ and $\frac{1}{2}\varepsilon_{\mu\alpha\nu\iota}R^{\alpha\nu\iota}\!=\!B_{\mu}$ for simplicity. Notice that (\ref{dep1}-\ref{dep2}) are equivalent to the initial Dirac spinor field equations, as it has been proven with detail in references \cite{Fabbri:2020ypd} and references therein.

Finally, from (\ref{dep1}-\ref{dep2}) we can derive
\begin{eqnarray}
&\!\!\!\!P^{\rho}\!=\!m\cos{\beta}u^{\rho}\!-\!Y\!\cdot\!s u^{\rho}
\!+\!Y\!\cdot\!u s^{\rho}\!+\!Z_{\mu}s_{\alpha}u_{\nu}\varepsilon^{\mu\alpha\nu\rho}
\label{momentum}
\end{eqnarray}
with $Y_{\mu}\!=\!\frac{1}{2}(\nabla_{\mu}\beta\!+\!B_{\mu})$ and $Z_{\mu}\!=\!-\frac{1}{2}(\nabla_{\mu}\ln{\phi^{2}}\!+\!R_{\mu})$ as explicit expression of the momentum \cite{Fabbri:2020ypd}.
\section{Plane-Wave Reduction}
\label{IV}
Having presented the polar re-formulation of the Dirac theory, we are next going to employ it to find in what way plane waves are contained in it. The structure of a plane-wave solution can be covariantly expressed according to
\begin{eqnarray}
&i\boldsymbol{\nabla}_{\mu}\psi\!=\!P_{\mu}\psi\label{pw}
\end{eqnarray}
where $P_{\mu}$ is the momentum. Such a condition can be thought as the position representation of the position-momentum commutation relations of quantum mechanics, in relativistic form, applied to spinor fields. When the assumption of plane waves is implemented, we have that (\ref{decspinder}) reduces to (\ref{pw}) only if
\begin{eqnarray}
&R_{ij\mu}\!=\!0\\
&\beta\!=\!0\\
&\nabla_{\mu}\phi\!=\!0
\end{eqnarray}
since a constant pseudo-scalar must be zero identically. In this case the field equations (\ref{dep1}-\ref{dep2}) become
\begin{eqnarray}
&P^{\iota}u_{[\iota}s_{\mu]}\!-\!s_{\mu}m\!=\!0\label{dep1pw}\\
&P^{\rho}u^{\nu}s^{\alpha}\varepsilon_{\mu\rho\nu\alpha}\!=\!0\label{dep2pw}
\end{eqnarray}
as easy to see. Then, the momentum (\ref{momentum}) reduces to
\begin{eqnarray}
&P^{\rho}\!=\!mu^{\rho}\label{momentumpw}
\end{eqnarray}
as expected. By plugging this back into (\ref{dep1pw}-\ref{dep2pw}) we see that these are verified identically. Thus in the case of plane waves, the Dirac equations are equivalent to the single condition $P^{\rho}\!=\!mu^{\rho}$ showing that the $8$ Dirac equations reduce to only $4$ independent constraints. This halving of independent components is well known, albeit it is usually thought to come from the on-shell condition, and not from the plane-wave condition. Here we have shown that it is the plane-wave condition what locks the two chiral parts to be equal, halving the degrees of freedom of spinors. Finally, consider the plane-wave condition applied to the spin-sum relations. In (\ref{Fierzpolarregular}) we can plug $\beta\!=\!0$ to get the much simpler
\begin{eqnarray}
&\psi\overline{\psi}\!\equiv\!\frac{1}{2}(m\mathbb{I}\!+\!P_{a}\boldsymbol{\gamma}^{a})
(\mathbb{I}\!-\!s_{a}\boldsymbol{\gamma}^{a}\boldsymbol{\pi})
\end{eqnarray}
having normalized $\phi^{2}\!=\!m$ and having used (\ref{momentumpw}) again. As it is possible to witness, these are precisely the Michel-Wightman identities \cite{mw}. In spin-average
\begin{eqnarray}
&\sum_{\mathrm{spin}}\psi\overline{\psi}\!\equiv\!(m\mathbb{I}\!+\!P_{a}\boldsymbol{\gamma}^{a})
\end{eqnarray}
which coincide with those we have in standard QFT.

Because the Michel-Wightman identities constitute a generalization of the common spin sums, the identities (\ref{Fierzpolarregular}) are yet another generalization. In fact, since they are obtained from (\ref{regular}), which is the most general form of spinor field, the identities (\ref{Fierzpolarregular}) are the most general form of spin sums that we can have.

With these spin sums one can now compute the propagators for internal lines, and so the scattering amplitudes. In the rest of the paper we will present, as an example, an application to a specific case.
\section{General Computation}
\label{V}
Despite aiming to work in general cases, one assumption that we shall still consider will be that of taking into account the spin averages. So (\ref{Fierzpolarregular}) reduces to
\begin{eqnarray}
\sum_{\mathrm{spin}}\psi\overline{\psi}\!=\!\frac{1}{p^{2}\!-\!m^{2}}
(p\!\!\!/\!+\!me^{-i\beta\boldsymbol{\pi}})\label{prop}
\end{eqnarray}
having defined $p^{a}\!=\!mu^{a}$ as the kinematic momentum in terms of mass and velocity and having used the same normalization commonly used in QFT. The above is the form of the propagator for fermionic internal lines.

It is important to notice that even in spin average the kinematic momentum $p_{\rho}$ is still different, if $\beta$ is different from zero, from the dynamic momentum $P_{\rho}$ in (\ref{momentum}). With the dynamic momentum instead, the spin sum would be
\begin{eqnarray}
\sum_{\mathrm{spin}}\psi\overline{\psi}\!=\!\frac{\cos{\beta}}{P^{2}\!-\!|m\cos{\beta}|^{2}}
(P\!\!\!\!/\!+\!m\cos{\beta}e^{-i\beta\boldsymbol{\pi}})
\end{eqnarray}
in terms of the cosine of the chiral angle.

For a small chiral angle, the two expressions coincide. In general, there are no compelling reasons to limit the investigation to the kinematic momentum, instead of the dynamic momentum. Here however, we will only consider the kinematic momentum, that is (\ref{prop}), since this is what is done in QFT.
\section{The Compton Scattering}
\label{VI}
It is now time to apply the above results to the actual computations in a specific case. Because we want to see the effects of the chiral angle $\beta$ in the propagator, spinors must constitute the internal lines. And since the $\beta$ angle is usually non-zero for interacting fermions, we ought to pick a case in which some fermion is in a bound state. The perfect example is the Compton scattering \cite{ps}. In this case the photon scatters off an electron that is bound around the nucleus, and consequently it has a non-zero chiral angle.

The invariant matrix element (\ref{M2}) is
\begin{eqnarray}
\mathcal{M}\!=\!-e^{2}\overline{u}(p')a_{\mu}(k')\boldsymbol{\gamma}^{\mu}
\frac{p\!\!\!/\!+\!k\!\!\!/\!+\!me^{-i\beta\boldsymbol{\pi}}}{2p\!\cdot\!k}
\boldsymbol{\gamma}^{\nu}a_{\nu}(k)u(p)
\end{eqnarray}
as $k^{2}\!=\!0$ and $p^{2}\!=\!m^{2}$ for photon and electron. The full treatment of the Compton scattering involves also the complementary diagram
\begin{eqnarray}
\mathcal{M}\!=\!+e^{2}\overline{u}(p')a_{\mu}(k')\boldsymbol{\gamma}^{\nu}
\frac{p\!\!\!/\!-\!k\!\!\!/'\!+\!me^{-i\beta\boldsymbol{\pi}}}{2p\!\cdot\!k'}
\boldsymbol{\gamma}^{\mu}a_{\nu}(k)u(p)
\end{eqnarray}
now with $k'^{2}\!=\!0$ for the photon. The full invariant matrix element is thus
\begin{eqnarray}
\mathcal{M}\!=\!-e^{2}\overline{u}(p')a_{\mu}(k')[\boldsymbol{\gamma}^{\mu}
\frac{p\!\!\!/\!+\!k\!\!\!/\!+\!me^{-i\beta\boldsymbol{\pi}}}{2p\!\cdot\!k}
\boldsymbol{\gamma}^{\nu}\!+\!\boldsymbol{\gamma}^{\nu}
\frac{-p\!\!\!/\!+\!k\!\!\!/'\!-\!me^{-i\beta\boldsymbol{\pi}}}{2p\!\cdot\!k'}
\boldsymbol{\gamma}^{\mu}]a_{\nu}(k)u(p)
\end{eqnarray}
which now must be squared. Taking the trace we get
\begin{eqnarray}
\nonumber
\frac{1}{4}|\mathcal{M}|^{2}\!=\!\frac{e^{4}}{4}
\mathrm{tr}\left[(p\!\!\!/'\!+\!m)[\boldsymbol{\gamma}^{\mu}
\frac{p\!\!\!/\!+\!k\!\!\!/\!+\!me^{-i\beta\boldsymbol{\pi}}}{2p\!\cdot\!k}
\boldsymbol{\gamma}^{\nu}\!+\!\boldsymbol{\gamma}^{\nu}
\frac{-p\!\!\!/\!+\!k\!\!\!/'\!-\!me^{-i\beta\boldsymbol{\pi}}}{2p\!\cdot\!k'}
\boldsymbol{\gamma}^{\mu}]\cdot\right.\\
\left.\cdot(p\!\!\!/\!+\!m)[\boldsymbol{\gamma}_{\nu}
\frac{p\!\!\!/\!+\!k\!\!\!/\!+\!me^{-i\beta\boldsymbol{\pi}}}{2p\!\cdot\!k}
\boldsymbol{\gamma}_{\mu}\!+\!\boldsymbol{\gamma}_{\mu}
\frac{-p\!\!\!/\!+\!k\!\!\!/'\!-\!me^{-i\beta\boldsymbol{\pi}}}{2p\!\cdot\!k'}
\boldsymbol{\gamma}_{\nu}]\right]
\end{eqnarray}
having used $u\overline{u}\!=\!p\!\!\!/\!+\!m$ and $a_{\sigma}a_{\nu}\!=\!-g_{\sigma\nu}$ for the external legs of both the electron and the photon contributions.

From now on all goes as standard QFT. Long but straightforward computations, involving Clifford algebras, and especially the traces of the Clifford matrices, lead to
\begin{widetext}
\begin{eqnarray}
\nonumber
\!\!\!\!\!\!\!\!\frac{1}{4}|\mathcal{M}|^{2}
\!=\!2e^{4}\left[\frac{\omega}{\omega'}\!+\!\frac{\omega'}{\omega}
\!+\!4m\left(\frac{1}{\omega}\!-\!\frac{1}{\omega'}\right)
\!+\!3m^{2}\left(\frac{1}{\omega'^{2}}\!+\!\frac{1}{\omega^{2}}\right)-\right.\\
\left.-2m\cos{\beta}\left[2m\left(\frac{1}{\omega\omega'}
\!+\!\frac{1}{\omega^{2}}\!+\!\frac{1}{\omega'^{2}}\right)
\!+\!\left(\frac{1}{\omega}\!-\!\frac{1}{\omega'}\right)\right]
\!+\!2m^{2}\cos{(2\beta)}\left(\frac{1}{\omega\omega'}
\!+\!\frac{1}{\omega^{2}}\!+\!\frac{1}{\omega'^{2}}\right)\right]\label{M}
\end{eqnarray}
\end{widetext}
since $p'\!\cdot\!k'\!=\!p\!\cdot\!k\!=\!m\omega$ and $k\!\cdot\!p'\!=\!p\!\cdot\!k'\!=\!m\omega'$ and we also have $p\!\cdot\!p'\!=\!k\!\cdot\!k'\!+\!m^{2}$ as well as $k\!\cdot\!k'\!=\!m(\omega\!-\!\omega')$ as usual.

As it is clear, the contribution of the chiral angle appears to modify the standard expression. For the first-order perturbative in $\beta$ it reduces to
\begin{widetext}
\begin{eqnarray}
\!\!\!\!\!\!\!\!\frac{1}{4}|\mathcal{M}|^{2}
\!\approx\!2e^{4}\left[\frac{\omega}{\omega'}\!+\!\frac{\omega'}{\omega}
\!-\!2m\left(\frac{1}{\omega'}\!-\!\frac{1}{\omega}\right)
\!+\!m^{2}\left(\frac{1}{\omega'}\!-\!\frac{1}{\omega}\right)^{2}
\!-\!\beta^{2}m^{2}\left[2\left(\frac{1}{\omega\omega'}
\!+\!\frac{1}{\omega^{2}}\!+\!\frac{1}{\omega'^{2}}\right)
\!+\!\frac{1}{m}\left(\frac{1}{\omega'}\!-\!\frac{1}{\omega}\right)\right]\right]\label{m}
\end{eqnarray}
\end{widetext}
where the standard expression is now obvious and it is accompanied by an expression that is quadratic in the chiral angle. The chiral angle $\beta$ appears squared as expected from the fact that it is a pseudo-scalar while $|\mathcal{M}|^{2}$ is a scalar.

To evaluate the correction, we recall that the scattered electron is in a bound state with its nucleus, and therefore it is natural to assume it is described by the $1S$ orbital of hydrogen-like atoms. In this case we know what is the form of the chiral angle, as it is given by
\begin{eqnarray}
\beta\!=\!-\arctan{\left(e^{2}\cos{\theta}/\sqrt{1\!-\!e^{4}}\right)}\!\approx\!-e^{2}\cos{\theta}
\end{eqnarray}
where $\theta$ is the elevation in spherical coordinates.\! The fact that it is point-dependent is not surprising for a function, but for better applicability it is more convenient to use a mean over all possible orientations, which is given by
\begin{eqnarray}
\beta^{2}_{\mathrm{mean}}
\!=\!\frac{1}{4\pi}\int \beta^{2}\sin{\theta}d\theta d\varphi
\!\approx\!\frac{e^{4}}{3}
\end{eqnarray}
of the same magnitude of $e^{2}$ itself.

Plugging it into (\ref{m}), and then plugging the result into the cross-section, furnishes
\begin{eqnarray}
\frac{d\sigma}{d\cos{\theta}}\!\approx\!\frac{\pi \alpha^{2}}{m^{2}}
\left(\frac{\omega'}{\omega}\right)^{2}\left[\frac{\omega}{\omega'}
\!+\!\frac{\omega'}{\omega}\!-\!|\!\sin{\theta}|^{2}
\!-\!\frac{16\pi^{2}\alpha^{2}}{3}\left[2m^{2}\!\left(\frac{1}{\omega\omega'}
\!+\!\frac{1}{\omega^{2}}\!+\!\frac{1}{\omega'^{2}}\right)
\!+\!1\!-\!\cos{\theta}\right]\right]
\end{eqnarray}
having used the Compton relation
\begin{eqnarray}
\frac{1}{\omega'}\!-\!\frac{1}{\omega}\!=\!\frac{1}{m}(1\!-\!\cos{\theta})\label{Compton}
\end{eqnarray}
and $e^{2}\!=\!4\pi\alpha$ for correction to the Klein-Nishina formula.

In this computation, the electron is in bound state after photon absorption.\! But we may also consider the electron in bound state before photon absorption. Then
\begin{eqnarray}
\sum_{\mathrm{spin}}\psi\overline{\psi}\!=\!\frac{\phi^{2}}{m}(me^{-i\beta\boldsymbol{\pi}}
\!+\!p_{a}\boldsymbol{\gamma}^{a})
\end{eqnarray}
is the propagator for external lines. Internal lines are now treated as describing electrons in the free state.

The invariant matrix element is assigned as above and when we compute its square, after taking the trace, we get
\begin{eqnarray}
\frac{1}{4}|\mathcal{M}|^{2}\!=\!\frac{\phi^{2}}{m}\frac{e^{4}}{4}
\mathrm{tr}\left[[\boldsymbol{\gamma}^{\mu}
\frac{p\!\!\!/\!+\!k\!\!\!/\!+\!m}{2p\!\cdot\!k}
\boldsymbol{\gamma}^{\nu}\!+\!\boldsymbol{\gamma}^{\nu}
\frac{-p\!\!\!/\!+\!k\!\!\!/'\!-\!m}{2p\!\cdot\!k'}
\boldsymbol{\gamma}^{\mu}](p\!\!\!/\!+\!me^{-i\beta\boldsymbol{\pi}})[\boldsymbol{\gamma}_{\nu}
\frac{p\!\!\!/\!+\!k\!\!\!/\!+\!m}{2p\!\cdot\!k}
\boldsymbol{\gamma}_{\mu}\!+\!\boldsymbol{\gamma}_{\mu}
\frac{-p\!\!\!/\!+\!k\!\!\!/'\!-\!m}{2p\!\cdot\!k'}
\boldsymbol{\gamma}_{\nu}](p\!\!\!/'\!+\!m)\right]
\end{eqnarray}
where the external leg for the outgoing electron is taken in plane waves as the electron leaving the nucleus is free.

Again, computations are now as usual, just a little easier because of the single appearance of the chiral angle in the full expression. After they are done we get
\begin{widetext}
\begin{eqnarray}
\frac{1}{4}|\mathcal{M}|^{2}\!=\!\frac{\phi^{2}}{m}2e^{4}\left[
\frac{\omega}{\omega'}\!+\!\frac{\omega'}{\omega}
\!-\!2m\left(\frac{1}{\omega'}\!-\!\frac{1}{\omega}\right)
\!+\!m^{2}\left(\frac{1}{\omega'}\!-\!\frac{1}{\omega}\right)^{2}
\!-\!(1-\cos{\beta})m^{2}\left[\left(\frac{2}{\omega^{2}}\!+\!\frac{2}{\omega'^{2}}\!-\!\frac{1}{\omega\omega'}\right)
\!-\!\frac{3}{m}\left(\frac{1}{\omega'}\!-\!\frac{1}{\omega}\right)\right]\right]\label{M'}
\end{eqnarray}
\end{widetext}
with the same notations used above.

In the Compton scattering the electron is in the bound state described by the $1S$ orbital, of which we already know the expression in polar form, and therefore we know how to evaluate the chiral angle. When solutions of the Coulomb potentials will provide the structure of $2S$ and $2P$ orbitals in polar form, we will be able to evaluate the chiral angle also in those cases, and the computation of the energy splitting in the Lamb shift will also become accessible.

Generally speaking, the general spin sum (\ref{Fierzpolarregular}) may always be relevant in cases in which the internal lines are fermionic. This is true in particular when radiative corrections are considered. In this case, the simplest process is the electron vertex
\begin{figure}[H]
\includegraphics[bb=10 570 600 780,scale=0.7]{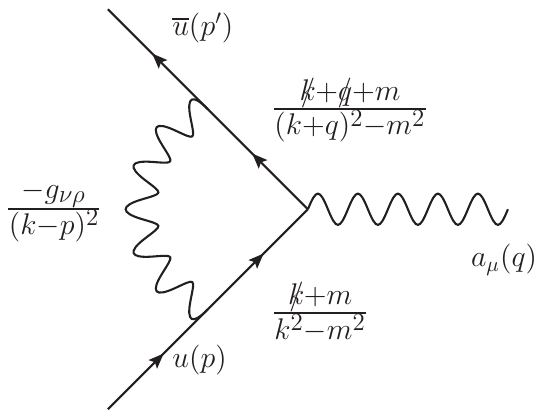}
\label{fig-3}
\end{figure}
\noindent resulting into the invariant matrix element
\begin{eqnarray}
\mathcal{M}\!=\!\frac{1}{(2\pi)^{4}}\int 
\overline{u}(p')(ie\boldsymbol{\gamma}^{\nu})
\frac{k\!\!\!/\!+\!q\!\!\!/\!+\!m\mathbb{I}}{(k\!+\!q)^{2}\!-\!m^{2}}
(ie\boldsymbol{\gamma}^{\mu})a_{\mu}(q)\frac{-g_{\nu\rho}}{(k\!-\!p)^{2}}
\frac{k\!\!\!/\!+\!m\mathbb{I}}{k^{2}\!-\!m^{2}}
(ie\boldsymbol{\gamma}^{\rho})u(p)d^{4}k
\end{eqnarray}
where we have to take the propagator as (\ref{prop}). After its substitution we obtain the expression
\begin{eqnarray}
\mathcal{M}\!=\!\frac{ie^{3}}{(2\pi)^{4}}\int 
\overline{u}(p')\boldsymbol{\gamma}^{\nu}
\frac{k\!\!\!/\!+\!q\!\!\!/\!+\!me^{-i\beta\boldsymbol{\pi}}}{(k\!+\!q)^{2}\!-\!m^{2}}
\boldsymbol{\gamma}^{\mu}a_{\mu}(q)\frac{1}{(k\!-\!p)^{2}}
\frac{k\!\!\!/\!+\!me^{-i\zeta\boldsymbol{\pi}}}{k^{2}\!-\!m^{2}}
\boldsymbol{\gamma}_{\nu}u(p)d^{4}k
\end{eqnarray}
where $\beta\!=\!\beta(k\!+\!q)$ and $\zeta\!=\!\zeta(k)$ in general.

The evaluation can be done again by neglecting the spin-dependent terms and it leads to the correction expressed in terms of two contributions as
\begin{widetext}
\begin{eqnarray}
\delta\mathcal{M}_{\mathrm{s}}\!=\!2ie^{3}\frac{-2m}{(2\pi)^{4}}\int \overline{u}(p')u(p)
\frac{k^{\mu}(1\!-\!\cos{\beta})\!+\!(k\!+\!q)^{\mu}(1\!-\!\cos{\zeta})}
{[(k\!+\!q)^{2}\!-\!m^{2}][k^{2}\!-\!m^{2}](k\!-\!p)^{2}}
a_{\mu}(q)d^{4}k
\end{eqnarray}
\begin{eqnarray}
\delta\mathcal{M}_{\mathrm{v}}\!=\!2ie^{3}\frac{m^{2}}{(2\pi)^{4}}\int 
\overline{u}(p')\boldsymbol{\gamma}^{\mu}u(p)
\frac{1\!-\!\cos{(\beta\!-\!\zeta)}}
{[(k\!+\!q)^{2}\!-\!m^{2}][k^{2}\!-\!m^{2}](k\!-\!p)^{2}}
a_{\mu}(q)d^{4}k
\end{eqnarray}
\end{widetext}
the first being scalar and the second being vectorial. For the specific case in which $\beta\!\approx\!\zeta$ the vector contribution would disappear leaving only the scalar contribution
\begin{widetext}
\begin{eqnarray}
\delta\mathcal{M}_{\mathrm{s}}\!=\!2ie^{3}\frac{-2m}{(2\pi)^{4}}\int \overline{u}(p')u(p)
(1\!-\!\cos{\beta})\frac{(2k\!+\!q)^{\mu}}
{[(k\!+\!q)^{2}\!-\!m^{2}][k^{2}\!-\!m^{2}](k\!-\!p)^{2}}a_{\mu}(q)d^{4}k
\end{eqnarray}
\end{widetext}
which can be evaluated following the common prescriptions of QFT. Nevertheless, from this point on, the problem we meet is that we do not have the expression of the chiral angle for electrons.\! We know that non-trivial expressions for the chiral angle can only be present when some internal dynamics is allowed for electrons, and thus within a more realistic description in which electrons are not treated as point particles. In lack of such a description we cannot go forward. However, we may still wonder what happens if the chiral angle were not dependent on the momentum of the particle. In this case
\begin{widetext}
\begin{eqnarray}
\delta\mathcal{M}_{\mathrm{s}}\!=\!-ie\frac{4me^{2}}{(2\pi)^{4}}(1\!-\!\cos{\beta})\!
\int \overline{u}(p')u(p)\frac{(2k\!+\!q)^{\mu}}
{[(k\!+\!q)^{2}\!-\!m^{2}][k^{2}\!-\!m^{2}](k\!-\!p)^{2}}a_{\mu}(q)d^{4}k
\end{eqnarray}
\end{widetext}
and in the approximation of vanishing momentum transfer $q\!\rightarrow\!0$ and eventually
\begin{widetext}
\begin{eqnarray}
\delta\mathcal{M}_{\mathrm{s}}\!=\!-iea_{\mu}(0)
\frac{16m^{2}e^{2}}{(2\pi)^{4}}(1\!-\!\cos{\beta})\!
\int \frac{k^{\mu}}{[k^{2}\!-\!m^{2}]^{2}(k\!-\!p)^{2}}d^{4}k
\end{eqnarray}
\end{widetext}
which can finally be evaluated. However, re-normalizability arguments would lead us away from our scope.

Another situation in which internal lines are relevant is in the instance of the electron self-energy. Nevertheless, corrections due to the chiral angle are present again only for electrons displaying some form of internal dynamics.

Very generally, any application of the generalized QFT formalism is based on the knowledge of the chiral angle, and so computations can be carried on only when non-trivial solutions of freely propagating electrons will be known.

The actual procedure to find such exact non-trivial solutions for freely-propagating spinor fields, however, is not at all an easy one. The search for these exact solutions is a subject that must be tackled with the latest tools of the polar decomposition of spinors \cite{Fabbri:2021weq}, still in its early stage. Perhaps more results will be available in the future.
\section{Conclusion}
In this work we have considered the problem of evaluating scattering amplitudes in terms of internal lines described by propagators written in terms of polarization-sum and spin-sum relationships. We have recalled that these spin sums are normally computed for the plane-wave solutions of free field equations. For fermions, such standard spin sums computed with plane waves can be seen as the spin-average approximation of enlarged spin sums called Michel-Wightman identities. We have seen that these Michel-Wightman spin sums are themselves a restricted case of yet more encompassing spin sums given by the identities (\ref{Fierzpolarregular}). These last spin sums are the most general we can have.

Quantum Field Theory itself, in its formal structure, was not modified, so the propagators and scattering amplitudes were the standard ones. We used the above general spin sums into the usual propagators to compute scattering amplitudes for the Compton effect as an example. Its purpose was to show that the evaluation of scattering amplitudes can be done for various processes exactly as in standard Quantum Field Theory.

\end{document}